\documentclass{article}

\usepackage[utf8]{inputenc}
\usepackage[a4paper, margin=1.2in]{geometry}
\usepackage{setspace}
\usepackage{amsmath, amsthm, amssymb}
\usepackage{graphicx,float}
\usepackage{tikz}
\usetikzlibrary{arrows.meta, positioning, shapes.geometric}
\usepackage{natbib}

\usepackage{amsmath,amsfonts,bm}

\def\1{\bm{1}}

\DeclareMathAlphabet{\mathsfit}{\encodingdefault}{\sfdefault}{m}{sl}
\SetMathAlphabet{\mathsfit}{bold}{\encodingdefault}{\sfdefault}{bx}{n}

\def\gI{{\mathcal{I}}}

\def\gN{{\mathcal{N}}}

\def\sA{{\mathbb{A}}}

\def\sI{{\mathbb{I}}}

\def\sS{{\mathbb{S}}}

\newcommand{\E}{\mathbb{E}}

\newcommand{\R}{\mathbb{R}}

\newcommand{\Var}{\mathrm{Var}}

\newtheorem{example}{Example}

\title{Adaptive Bayesian Learning with Action and State-Dependent Signal Variance}
\author{Kaiwen Hou\thanks{KHou24@gsb.columbia.edu}\\Columbia Business School}
\date{}

\begin{document}

\maketitle
\begin{abstract}
    This manuscript presents an advanced framework for Bayesian learning by incorporating action and state-dependent signal variances into decision-making models. This framework is pivotal in understanding complex data-feedback loops and decision-making processes in various economic systems. Through a series of examples, we demonstrate the versatility of this approach in different contexts, ranging from simple Bayesian updating in stable environments to complex models involving social learning and state-dependent uncertainties. The paper uniquely contributes to the understanding of the nuanced interplay between data, actions, outcomes, and the inherent uncertainty in economic models.
\end{abstract}
\section{Introduction}
Bayesian learning, a fundamental concept in statistical inference and decision-making, has gained significant traction across various fields due to its ability to integrate prior knowledge with new information. As a robust methodology, Bayesian learning has been widely acknowledged for its adaptability and precision in handling uncertainty and updating beliefs~\citep{gelman1995bayesian}. 

This manuscript expands upon the Bayesian learning framework~\citep{baley2023bayesian} through uniquely addressing the action and state-dependent signal variance in the agents' information set. 
At the core of this framework is the concept that the precision of the signal received by an agent is contingent upon both the agent's action and the actual state, for example, based on their congruence or tracking error~\citep{daly2018feasible,du2021comparing,orlik2014understanding,rompotis2011predictable,stone2013bayesian,yang2022active}. For instance, when an action perfectly aligns with the true state, the agent is rewarded with a signal of perfect precision, devoid of any variance. Conversely, any deviation between the action and the true state results in increased signal variance, thus introducing greater uncertainty into the agent's decision-making process.

This perspective is crucial for a more sophisticated and detailed understanding of decision-making in dynamic environments. Additionally, the adaptive Bayesian learning framework presented in this paper provides an enhanced insight into the data-feedback loop, as conceptualized by \citet{farboodi2021model}. By accommodating a broader range of sophisticated scenarios within the data economy, this framework elucidates the intricate dynamics of the feedback loop, emphasizing its prevalence and impact in various contexts. This approach not only expands upon traditional Bayesian models but also contributes to a deeper comprehension of the interplay between data, actions, and outcomes in complex economic systems.

We delve into various scenarios within the data-driven economic landscape to showcase how this framework can serve as a unifying approach to understanding the nuanced interplay between actions, states, and signal precision. From simple Bayesian updating in stable environments to intricate models of social learning and state-dependent uncertainties, our exploration reveals the versatility and applicability of this framework across multiple domains. 

\section{Problem Formalization}

\subsection{State}
The state of the world, denoted as \( a^* \in \sS \subset\R \), is an unknown and dynamic element that the agent seeks to estimate. Following a common approach in Bayesian inference due to the distribution's mathematical properties and its ability to model a wide range of scenarios, the agent's prior belief about $a^*$ is modeled as a normal distribution
$$
 a^* \sim \gN(\mu_a, \sigma_a^2).
$$

\subsection{Action}
The agent chooses an action \( a\in\sA \subset \R\) based on their current belief about the state \( a^* \). This highlights the decision-making aspect of the agent, where actions are taken based on their understanding and estimation of the state.

\subsection{Signal}
The signal $s$ observed by the agent is a crucial aspect of the model. It is affected by noise $\epsilon$, which is assumed to be normally distributed with a variance that depends on both the true state and the agent's action. This dependency from traditional Bayesian learning frameworks is critical, as it introduces a direct link between the agent's actions and the uncertainty in the information they receive. More formally, the observed signal is given by 
$$
s = a^* + \epsilon, \quad \epsilon \sim \gN\left(0,\Sigma_\epsilon(a^*,a)\right),
$$
with a noise variance function $\Sigma_\epsilon:\sS\times\sA\to\R$.

\subsection{Bayesian Updating}
Upon receiving signal $s$, the agent revises their belief about state $a^*$. This revision is affected by the action selected, due to the variance of noise that depends on the action. More precisely, the posterior distribution \( p(a^* | s, a) \), as opposed to merely \( p(a^* | s) \), is influenced by the selected action $a$, factoring in the action-dependent noise variance. This underscores the dynamic nature of the agent's belief system, where each new piece of information has the potential to modify their perception of the state.

\section{Examples}
In this section, we explore different scenarios within the data economy where the concept of action-state-dependent signal precision can be applied as a unifying framework.

\begin{example}[Simple Bayesian Updating~\citep{joyce2003bayes}]
Consider a scenario where $\Sigma_\epsilon(a^*,a)\equiv\sigma_\epsilon^2$ is constant. The posterior distribution then becomes:
\begin{equation}\label{eq:posterior_simple}
    a^*\mid s \sim \gN\left( \frac{\sigma_a^{-2}\mu_a+\sigma_\epsilon^{-2}s}{\sigma_a^{-2} + \sigma_\epsilon^{-2}}, 
\frac{1}{\sigma_a^{-2} + \sigma_\epsilon^{-2}} \right).
\end{equation}
\end{example}

This simplistic example illustrates a situation akin to an investor updating their beliefs about the state, such as the future performance of an asset, based on new information. The constant variance $\sigma_\epsilon^2$ suggests a stable environment, where the uncertainty about future returns does not change with new actions or information. For examples in this simple setting, refer to chapters 2-5 of \citet{veldkamp2023information}.

\subsection{Action-Dependent Signal Variance}
In the following, we present two examples, in which the uncertainty in the signal varies as the magnitude of the action $a$ varies. A concept here is commonly known as ``active learning" or ``active experimentation." For instance, it is exemplified in the bandit problem discussed in Section 3.2 of \citet{veldkamp2023information}, as well as in other examples found in \citet{bergemann2006bandit}. This approach contrasts with ``passive learning."

\begin{example}[Optimal Amount of Data]
The agent decides on $a$, the number of unbiased signals $s_i$ to gather, each with iid noises $\epsilon_i\sim\gN(0,\sigma_\epsilon^2)$. They form a sufficient statistic:
$$
s = \frac{1}{a} \sum_{i=1}^n s_i = a^* + \frac{1}{a} \sum_{i=1}^n \epsilon_i \triangleq a^* + \epsilon.
$$
The noise variance then depends on the number of signals:
$$
\Sigma_\epsilon(a^*,a) = \frac{\sigma_\epsilon^2}{a}.
$$
Similar to the posterior update in \eqref{eq:posterior_simple}, we have
$$
a^*\mid s,a \sim \gN\left( \frac{\sigma_a^{-2}\mu_a+a\sigma_\epsilon^{-2}s}{\sigma_a^{-2} + a\sigma_\epsilon^{-2}}, 
\frac{1}{\sigma_a^{-2} + a\sigma_\epsilon^{-2}} \right).
$$
\end{example}

This example is analogous to a firm making 
data-sensitive decisions, such as deciding what data to collect when conducting market research before launching a new product. The more data points (signals) the firm gathers, the lower the uncertainty in noise variance. Despite such a data-collection story, there might be a more sophisticated and heterogeneous underlying mechanism from which the ``amount" of data points is determined as shown in the following example. 

\begin{example}[Social Learning]
    In this model, we explore how interactions among agents $i$ lead to varied signal structures. Specifically, the noise variance can be represented as:
    $$
\Sigma_\epsilon(a^*,a) = \frac{\sigma_\epsilon^2}{\int_0^1 \sI_{a_i\in\sA_i}di},
    $$
    which reflects the diversity of actions within a potentially endogenous range $\sA_i\subset\sA$. Moreover, the posterior is
$$
a^*\mid s,a \sim \gN\left( \frac{\sigma_a^{-2}\mu_a+\sigma_\epsilon^{-2}s\int_0^1 \sI_{a_i\in\sA_i}di}{\sigma_a^{-2} + \sigma_\epsilon^{-2}\int_0^1 \sI_{a_i\in\sA_i}di}, 
\frac{1}{\sigma_a^{-2} + \sigma_\epsilon^{-2}\int_0^1 \sI_{a_i\in\sA_i}di} \right).
$$
\end{example}
The above example models the process of agents acquiring information through social learning, a concept derived from the understanding that firms learn from one another about various factors impacting their revenues, such as productivity, market demand, and regulatory changes~\citep{bikhchandani1998learning}. In this context, when an agent $i$ opts for a specific action $a=1$ (from the binary set $\{0,1\}$), like investing in a project, it generates a noisy signal about the project's return. This signal is then received by other agents, effectively becoming a public signal. The associated noise variance is contingent on the proportion of agents who chose action 1, leading to the expression:
$$
\Sigma_\epsilon(a^*,a) = \frac{\sigma_\epsilon^2}{\int_0^1 \sI_{a_i=1}di}.
$$
This formulation is vital for modeling ``uncertainty traps" as discussed in works by \citet{fajgelbaum2017uncertainty, saijo2017uncertainty}. The posterior mean $\hat\mu_a=\E[a^*\mid s,a]$ and variance $\hat\sigma_a^2=\Var[a^*\mid s,a]$ in these scenarios is updated according to the following:
$$
\hat\mu_{a,t+1}\mid \hat\mu_{at}, \hat\sigma_{at}^2, a_t \sim \gN\left(
\rho\hat\mu_{at}, \rho^2 \left( \frac{1}{\hat\sigma_{at}^{-2}}-\frac{1}{\hat\sigma_{at}^{-2}+\sigma_\epsilon^{-2}\int_0^1 \sI_{a_{it}=1}di}\right)
\right),
$$
where $\rho<1$ is an AR(1) coefficient in the state evolution of $a_t^*$. 

Most notably, this approach allows for the modeling of data as a by-product of economic activities. During periods of high activity (i.e., when $\int_0^1 \mathbb{I}_{a_i=1}di$ is large), more information is disseminated, leading to potentially greater fluctuations in beliefs. The model hence formally explores that agents are hesitant to make decisions due to uncertainty about the environment or the actions of others, which can lead to a lack of economic activities, further generating less data~\citep{kozeniauskas2018uncertainty} and exacerbating the uncertainty. As we can see, such a feedback loop between data and economic activities also amplifies the business cycle, where booms are times of high activities and information production~\citep{ordonez2013asymmetric,van2006learning,veldkamp2005slow}.

\subsection{State-Dependent Signal Variance}
It should also be noted that the uncertainty might be considered as a hidden state, as suggested by $a^*$, in alignment with the perspectives of \citet{ang2002regime} and \citet{orlik2014understanding}. In such scenarios, the functional form of noise variance could depend on the actual state, represented by $a^*$, as illustrated in the subsequent examples. This is quite realistic in financial markets, where periods of high volatility are often associated with greater unpredictability and more significant discrepancies between the signal and the true state. 
\begin{example}[Uncertainty Feedback]\label{ex:feedback}
    Consider a scenario where the state $a^*$ represents the volatility of the environment. Define the noise variance as $\Sigma_\epsilon(a^*,a) = f(a^*) \geq 0$. Under these conditions,
    $$
s\mid a^* \sim \gN\left(a^*,f(a^*)\right),
    $$
    and the posterior distribution is given by\footnote{Even in cases where the posterior does not align with a commonly used tractable distribution, numerical methods such as Markov Chain Monte Carlo (MCMC) can be effectively employed to sample from the distribution.}
    \begin{align*}
p(a^*\mid s) &\propto p(s\mid a^*)p(a^*) \\
&\propto f(a^*)^{-\frac{1}{2}} \exp\left( -\frac{(s-a^*)^2}{2f(a^*)} -\frac{(a^*-\mu_a)^2}{2\sigma_a^2}  \right).
    \end{align*}
    Figure~\ref{fig:uncertaintyFeedback} displays the normalized posterior density for different signal values.
    \begin{figure}[H]
        \centering
        \includegraphics[width=1.1\textwidth]{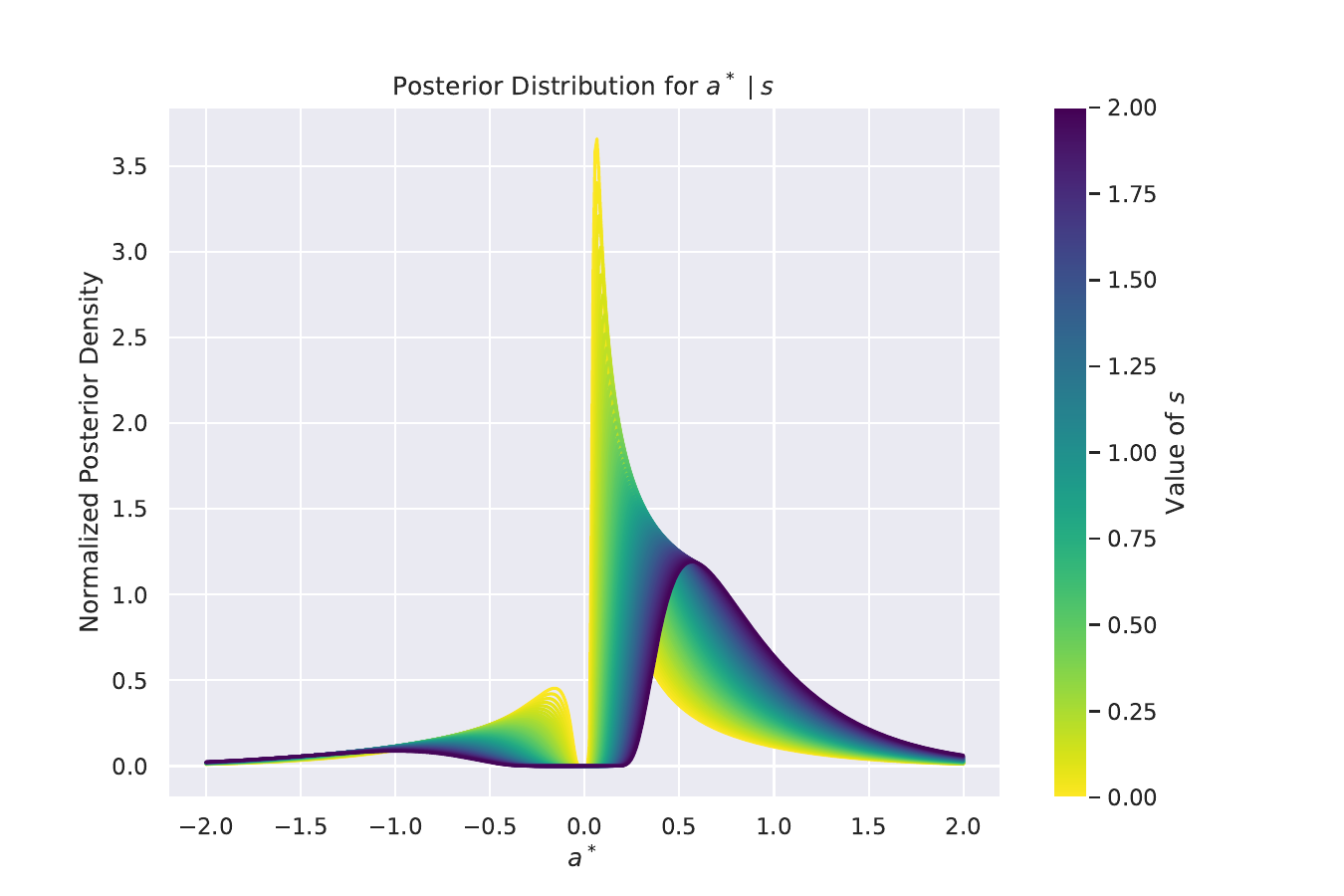}
        \caption{Posterior density of uncertainty feedback, with a standard normal prior, noise variance $f(a^*)=a^{*2}$, and different signal values.}
        \label{fig:uncertaintyFeedback}
    \end{figure}
\end{example}
In this scenario, the noise variance is conceptualized as a function of the actual state of volatility, denoted by $a^*$. This formulation models the relationship between volatility and the volatility of volatility. Notably, there is a positive correlation between VVIX and VIX, with several shared peaks, particularly during the financial crisis, as highlighted by \citet{huang2019volatility}. Additionally, the interdependence between the signal variance of volatility and the volatility itself is often influenced by volatility clustering. This phenomenon is characterized by sequences of high (or low) volatility days~\citep{ding1996modeling,granger1995some}. Furthermore, when stock returns are interpreted as a (potentially biased) signal of volatility, this relationship is exemplified by the leverage effect, which posits an inverse relationship between stock prices and their volatility~\citep{black1976studies,christie1982stochastic,schwert1989does}.

This formulation has also been used in, e.g., \citet{orlik2014understanding}. Consider a situation where the state \( a^* \) takes discrete values, \( a^*_1, a^*_2, \ldots, a^*_n \), each representing a distinct regime or state of the world. The agent has a prior belief about which state the world is in, represented by a discrete probability distribution over these states. Denote the probability of being in state \( a^*_i \) as \( p(a^* = a^*_i) \).

\begin{example}[Discrete Regimes]
The noise variance in this scenario is a function of the state, \( \Sigma_\epsilon(a^*,a) = f(a^*) \geq 0 \), where \( f \) is a function mapping states to their respective noise variances. This variance could represent the level of uncertainty or instability associated with each state.
The observed signal \( s \) is still given by 
$$
s\mid a^* \sim \gN\left(a^*,f(a^*)\right).
$$
Upon observing \( s \), the agent updates their belief about the current state \( a^* \) according to Bayes' theorem:
$$
p(a^* = a^*_i \mid s) = \frac{p(s \mid a^* = a^*_i) p(a^* = a^*_i)}{\sum_{j=1}^n p(s \mid a^* = a^*_j) p(a^* = a^*_j)}.
$$
\end{example}

Furthermore, in a regime-switching model~\citep{ang2002regime,bekaert2001peso,hamilton1989new}, if \( a^* \) evolves according to a Markov process, the transition probabilities between states can be used to update the prior probabilities \( p(a^* = a^*_i) \) for the next time period, based on the current posterior probabilities.
This model can be particularly useful in scenarios where different regimes or states have distinct characteristics and levels of uncertainty, such as in financial markets where different market conditions (e.g., bull market, bear market, or high volatility period) can significantly affect investment decisions and risk assessments.

\subsection{Action-State-Dependent Signal Variance}
\begin{example}[Tracking Error]\label{ex:tracking}
The signal noise variance can be related to the discrepancy between the action \( a \) and the true state \( a^* \):
$$
\Sigma_\epsilon(a^*,a) = k |a - a^*|^2 $$ for some constant \( k>0 \).
\end{example}

This model could represent a bandit problem, in which a firm implements a project $a_j$ and observes a project-specific output as a signal
$$
s_{j} =  A  -   (a_j - a_j^*)^2.
$$
The signal structure implies a constant output $A$ delivered by the implementation, minus a penalty for not being close to an unobserved project-specific target $a_j^*$.

Remarkably, \textbf{Example~\ref{ex:tracking}} extends to dynamic settings, linking forecast error to economic uncertainty~\citep{orlik2014understanding}. Here, agent $i$ forecasts the future state at $t+1$ using current information $\gI_{it}(a_{it})$ dependent on their action $a_{it}$. The ex-post forecast error at time $t$ is:
$$
\epsilon_{i,t+1} = \left| a_{t+1}^* - \E[a_{t+1}^*\mid \gI_{it}(a_{it})] \right|.
$$
In scenarios like predicting GDP growth, the environment's uncertainty is indicated by the expected forecast error. This could also lead to noisier public signals through the variance term:
$$
\Sigma_\epsilon(a_t^*,a_t) = \int_0^1 \E[\epsilon_{i,t+1}^2\mid \gI_{it}(a_{it})] di.
$$


\section{Conclusions}
Overall, this manuscript provides a comprehensive and nuanced understanding of Bayesian learning in dynamic environments, emphasizing the interplay between actions, states, and signal precision. The variety of examples presented showcases the framework's applicability across different scenarios, from stable environments to more complex situations involving heterogenous cross-sectional interactions and richer longitudinal structures.

\newpage
\onehalfspacing
\bibliographystyle{apalike}
\bibliography{literature.bib}

\end{document}